\documentclass[prl,twocolumn,showpacs,preprintnumbers]{revtex4}

\usepackage{graphicx,array}
\newcolumntype{C}[1]{>{\centering\let\newline\\\arraybackslash\hspace{0pt}}m{#1}}

\begin{document}
\preprint{}

\title{
Universal metastability of the low-spin state in Co$^{2+}$ systems: \\
non-Mott type pressure-induced spin-state transition in CoCl$_{2}$}




\author {Bongjae Kim, Kyoo Kim, and B. I. Min}
\affiliation{Department of Physics, PCTP,
Pohang University of Science and Technology, Pohang 790-784, Korea}
\date{\today}

\begin{abstract}
We have investigated the pressure-induced spin-state transition
in Co$^{2+}$ systems
in terms of a competition between the Hund's exchange energy ($J$)
and the crystal-field splitting ($\Delta_{CF}$).
First, we show the universal metastability of the low-spin state
in octahedrally coordinated Co$^{2+}$ systems.
Then we present the strategy to search for a Co$^{2+}$ system,
for which the mechanism of spin-state and metal-insulator transitions
is governed not by the Mott physics but by $J$ vs. $\Delta_{CF}$ physics.
Using CoCl$_{2}$ as a prototypical Co$^{2+}$ system,
we have demonstrated the pressure-induced spin-state transition
from high-spin to low-spin,
which is accompanied with insulator-to-metal and antiferromagnetic
to half-metallic ferromagnetic transitions.
Combined with metastable character of Co$^{2+}$ and the high compressibility
nature of CoCl$_{2}$, the transition pressure as low as 27 GPa can be
identified on the basis of $J$ vs. $\Delta_{CF}$ physics.
\end{abstract}


\maketitle



The spin-state transition in transition metal (TM) complexes
has been a subject of intense study.
The stabilization
of one spin-state over another is determined
by the competition of various energy scales,
such as Coulomb correlation ($U$), band-width ($W$),
Hund's exchange energy ($J$), and crystal-field splitting ($\Delta_{CF}$),
and so on.
Pressure induces the  spin-state transition by changing the relative strength
of different energy scales,
especially either $W$ or $\Delta_{CF}$.
In many cases, the pressure-induced spin-state
transition is accompanied by the metal-insulator transition,
which can be well described with either $J/\Delta_{CF}$ or $U/W$ ratio.
In real systems, both are important and interconnected.
The physics of $J$ vs. $\Delta_{CF}$ is hard
to explore in usual pressure study due to the entrance of $U$ vs. $W$ physics
in the form of $d$-$p$ hybridization or crystal distortion.
In a simple TM monoxide, MnO for example, as pressure increases,
the change of $\Delta_{CF}$ acts as primary role in the high-spin (HS)
to low-spin (LS) state transition and the insulator-to-metal transition.
However, the underlying physics is governed not only by $J$ and $\Delta_{CF}$
but also $U$ and $W$, and even by the charge transfer energy
($\Delta_{CT}=\varepsilon_{d}-\varepsilon_{p}$), which all take part
in the process of the spin state transition\cite{kunes08}.

 Studies of spin-state transition in Co-containing complexes have been mostly
concentrated on the Co$^{3+}$ ($d^6$) systems.
Well-known example is perovskite LaCoO$_{3}$.
LaCoO$_{3}$ shows temperature dependent spin-state transition,
which can be interpreted as LS ($S=0$) to HS ($S=2$) or
LS to intermediate-spin (IS) ($S=1$) transition.
The exact magnetic phase and the underlying mechanism are, however, still
under debate \cite{heikes,raccah,korotin,kunes11,krapek}.
Since the energy scales $J$ and $\Delta_{CF}$ in LaCoO$_{3}$ are similar,
thermal excitation can easily mix or switch
different spin-states in cooperation with lattice distortion \cite{raccah}.
Recently, Kune\v{s} \emph{et al.}\cite{kunes11} argued
the spin-state transition in terms of a
purely electronic origin without lattice effect.
Also, for LaCoO$_{3}$, it is known that pressure produces
the similar effect to temperature,
which is described as the depopulation of IS state
rather than phase change \cite{nekrasov,vogt,knizek,vanko}.

In contrast to Co$^{3+}$ systems,
the spin-state transition in Co$^{2+}$ ($d^7$)
systems is relatively unexplored.
It is because most of Co$^{2+}$ systems
have stable HS ($S=3/2$) states.
Nevertheless, the HS-LS transitions have been
discussed for a few Co$^{2+}$ systems, such as organic
complexes \cite{Jain75,Schmiedekamp02,Goodwin04,gutlich,Varadwaj10},
YBaCo$_{2}$O$_{5}$ \cite{Vogt00,kwon},
CoCl$_{2}$ \cite{hernandez},
and Ca$_{3}$Co$_{2-x}$Mn$_{x}$O$_{6}$ \cite{Flint10}.
Especially, for CoCl$_{2}$, the pressure-induced metallization
driven by the spin-state transition was indicated by
carrying out high-pressure optical absorption measurements \cite{hernandez}.
But the isostructural spin-state transition in Co$^{2+}$ systems
has not been confirmed experimentally and theoretically yet.

Motivated by the above investigations of
spin-state transition for Co$^{3+}$ systems,
we have studied energetics of
different spin-states of various octahedrally coordinated
Co$^{2+}$ (CoX$_{6}$) systems
on the basis of the {\it ab initio} electronic structure calculations.
Their magnetic properties are described by Co$^{2+}$ ions,
which have HS $3d^{7}$ (t$_{2g}^{5}e_{g}^{2}$) configurations
in the ground state.
Interestingly, all the tested systems have LS ($S=1/2$) metastable
states, and moreover their HS-LS energy differences are
of almost the same scale, independent of the
anion (X) type, Co-X bond length, and CoX$_{6}$ octahedron distortions.
We then discuss the strategy to search for the materials having
the pressure-driven spin-state transition governed by
$J$ vs. $\Delta_{CF}$ physics, unlike other TM oxides that show
the Mott-type transition governed by $U$ and $W$.
We propose that CoCl$_{2}$ is a prototypical Co$^{2+}$ system
having $J$ vs. $\Delta_{CF}$ physics. We show that CoCl$_{2}$ has
abrupt collapses in volume and spin magnetic moment
at the spin-state transition point, which is also
accompanied by the insulator-to-metal transition.


In order to investigate the spin-state transition in Co$^{2+}$ systems,
we have performed the electronic structure calculations employing
the full-potential linearized augmented plane wave
(FLAPW) band method \cite{Freeman} implemented in WIEN2k ~\cite{wien2k}.
Since the often-used pseudopotential band method
is known to have problems in describing the spin-states
under volume reduction,
the application of full-potential band method is essential
in the pressure studies~\cite{kolorenc}.
For the exchange-correlation energy functional,
we used the generalized gradient approximation (GGA)
with the PBEsol functional \cite{perdew1}. On-site Coulomb correlation is
treated with the GGA$+U$ method
in the rotationally invariant form ~\cite{anisimov,liechtenstein}.
The spin-orbit coupling (SOC) was included in the second variational scheme,
when necessary.

\begin{figure}[t]
\includegraphics[angle=270,width=76mm]{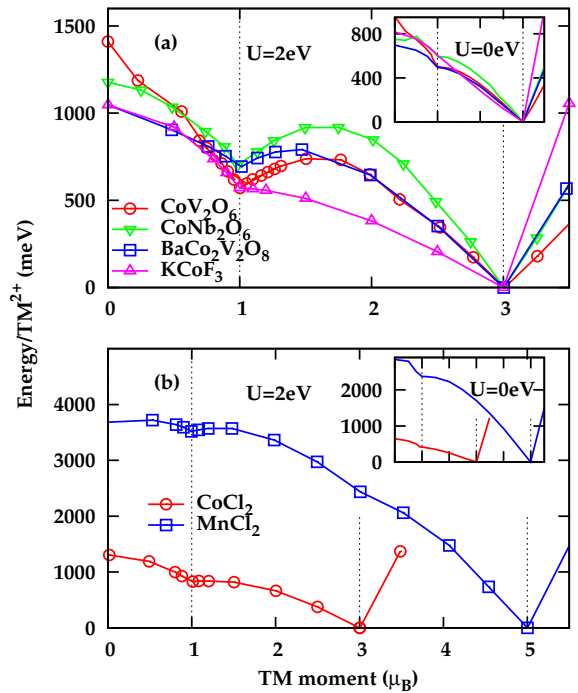}
\caption{(Color online)
Energy vs. spin magnetic moment obtained by using the FSM calculation.
(a) For octahedrally coordinated Co$^{2+}$ systems:
$\alpha$-CoV$_2$O$_6$, CoNb$_2$O$_6$,
BaCo$_2$V$_2$O$_8$, and KCoF$_3$.
(b) For two dichlorides: CoCl$_2$ and MnCl$_2$.
Results in the GGA+$U$ with $U$=2 eV are shown. Insets are for $U=0$.
Spin magnetic moment is obtained by dividing total magnetic moment
by number of Co ions
in the system to include the induced magnetic moment.
}
\label{fsm_systems}
\end{figure}

We first identify the general metastability
of LS phase in the octahedrally coordinated Co$^{2+}$ systems.
We have carried out the fixed-spin moment (FSM) calculations
for various Co$^{2+}$ systems, including
chain-type brannerite $\alpha$-CoV$_{2}$O$_{6}$ \cite{he09}
and CoNb$_{2}$O$_{6}$ \cite{heid},
complex chain-tetragonal BaCo$_{2}$V$_{2}$O$_{8}$ \cite{he05},
cubic perovskite-type cobaltate fluoride KCoF$_{3}$ \cite{kijima},
and simple layered rhombohedral cobaltate dichloride CoCl$_{2}$
\cite{hutchings}. All the above systems have Co as the only magnetic ion.
Figure~\ref{fsm_systems}(a) shows the energy vs. spin
magnetic moment for those Co$^{2+}$ systems.
Besides the global ground HS states at the moment ($M$) of
$3 \mu_{B}$/Co$^{2+}$, we can clearly see the metastable LS states
at $M= 1 \mu_{B}$/Co$^{2+}$ for all cases.
Furthermore, the HS-LS energy differences ($\Delta E$'s) are almost the same.
$\Delta E$'s are 400$-$600 meV and 570$-$830 meV for $U=0$ and
$U=2.0$ eV, respectively.
The inclusion of $U$ tends to increase $\Delta E$,
because the HS state is more favored with larger $U$.
However, the universal metastability of the LS phase remains the same.

The similarity in the energy scale of $\Delta E$ for different systems
is striking because not only the macroscopic
crystal structure but also the local CoX$_{6}$ (X=O, F and Cl)
octahedron structure are quite different from system to system.
For example, the Co-X bond length is very short for KCoF$_{3}$
(2.03{\AA}), but very  long for CoCl$_{2}$ (2.42{\AA})
even though both have almost ideal octahedra.
CoNb$_{2}$O$_{6}$, and BaCo$_{2}$V$_{2}$O$_{8}$
have the tetragonal distortions, and $\alpha$-CoV$_{2}$O$_{6}$ has
the additional in-plane rectangular distortion
that produces extraordinary crystal field levels \cite{kim}.
This feature suggests that the metastability of LS phase
is robust in the octahedrally coordinated Co$^{2+}$ systems.

In a simplified picture,
the energy difference between the LS (t$_{2g}^{6}e_{g}^{1}$)
and the HS (t$_{2g}^{5}e_{g}^{2}$) state
can be expressed as
\begin{equation}
\label {eq:diff}
\ \Delta E= E_{LS}-E_{HS}= 2J - \Delta_{CF}.
\end{equation}
Because the Hund's exchange $J$ does not have much system dependence,
the similar $\Delta E$ value for all systems suggests that $\Delta_{CF}$
does not vary much either in usual CoX$_{6}$ systems
despite the different coordinating structures.
To highlight the uniqueness of LS tendency in Co$^{2+}$ systems,
we have compared the LS-HS energetics of isostructural CoCl$_{2}$ and MnCl$_{2}$ systems.
Following the same idea of Eq. (1), Mn$^{2+}$ ($d^{5}$) has
$\Delta E= E_{LS}-E_{HS}= 6J - 2\Delta_{CF}$, and so
much larger LS-HS energy difference (by $4J - \Delta_{CF}$)
can be expected.
Indeed, as shown in Fig.~\ref{fsm_systems}(b), $\Delta E$ for MnCl$_{2}$
is obtained to be more than 2000 meV without $U$
(3500 meV with $U=2.0$ eV),
which is much larger than that for CoCl$_{2}$.

 Let us now consider the possible manipulation to make the metastable
LS phase be stabilized over the HS phase.

The easiest way is to apply the pressure.
When the pressure is applied on the system, $\Delta_{CF}$
will be increased due to
the contraction of Co-X distances, and eventually the HS to LS transition
can be realized when $\Delta_{CF}$ becomes larger than $J$.
However, as mentioned earlier,
the emergence of the Mott physics coming from
the change of $U$ and $W$
under pressure can complicate the approach.
It is worth noticing here that, according
to recent theoretical studies\cite{ovchinnikov,lyubutin},
the effective $U$ value for d$^{7}$ system is independent of the pressure,
while that for d$^{5}$ Mott system decreases under pressure.
Then we can safely assume the constant $U$ in the following pressure
studies.

To treat the Co$^{2+}$ system in term of $J$ vs. $\Delta_{CF}$ physics,
we have to consider two interconnected factors: $W$
and compressibility.
In the case of the TM oxides (fluorides), the hybridization
between TM-$3d$ and O-$2p$ (F-$2p$) bands
is very strong, and highly dependent on the pressure.
Hence, as the pressure increases, $W$ changes much and the resulting resilience
of the TM-O bond changes the shapes of octahedra too.
Thereby, pressure changes not only $\Delta_{CF}$ but also $W$.
Moreover, due to the rigidity of TM-$3d$ and O-$2p$ (F-$2p$) bond,
the structural transitions easily occur under pressure.
So there are overall changes in the octahedral rotation patterns and the
connectivity, which makes the system more dependent on $W$ value.

The compressibility, the volume change with respect to pressure,
should also be taken into account.
A system with high compressibility can be a good candidate
of the spin-state transition, since $\Delta_{CF}$
can be controlled under small pressure
without distorting the system or changing $W$ value.
It is known that chlorides usually have better
compressibility than oxides and fluorides \cite{aguado04},
because the hybridization of $d$-$p$ bands does not change much
upon pressure.

Based on the above criteria, we present CoCl$_{2}$
as an ideal model system
to study $J$ vs. $\Delta_{CF}$ physics for the following reasons:
(i) $W$ effect is much reduced due to weaker bonding
between Co-$d$ and Cl-$p$,
and accordingly the compressibility is very large.
(ii) No structural phase transition occurs up to the  spin-state
transition pressure \cite{hernandez}.
This is in contrast to CoF$_{2}$, for which
the structural transition is easily induced under pressure
even below 10 GPa \cite{ming}.
(iii) Not like other Co$^{2+}$ systems, the orbital magnetic moment is
small ($\sim10\%$ of the spin magnetic moment),
so that the magnetic behavior can be described
by the spin magnetic moment only \cite{wilkinson}.
(iv) The Jahn-Teller distortion is suppressed in the LS phase
\cite{hernandez}.
This is important because the Jahn-Teller distortion in many
half-filled e$_{g}$ orbital systems makes the situation
complicated \cite{choi}.
(v) Finally, the extensive band calculations are tractable
due to its rather simple structure (trigonal space group $R\bar{3}m$).

\begin{figure}[t]
\includegraphics[angle=270,width=76mm]{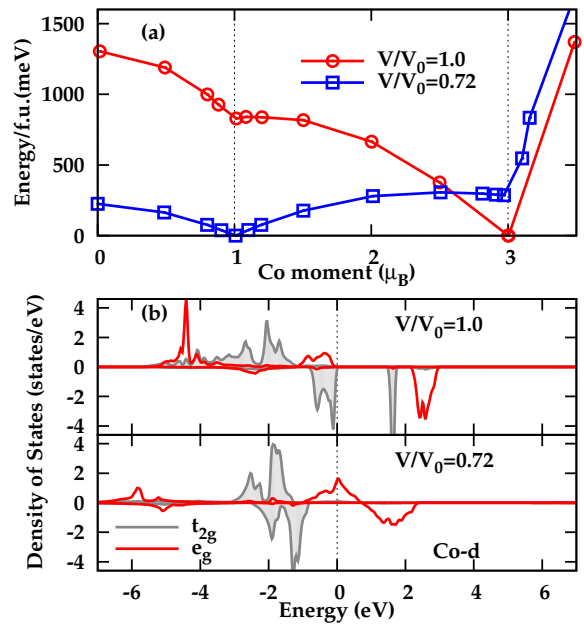}
\caption{(Color online) (a) The FSM calculations for CoCl$_{2}$
at the equilibrium volume
(V/V$_{0}$=1.0) and at the LS stabilized volume (V/V$_{0}$=0.72).
(b) PDOS of Co-$3d$ ($t_{2g}$ and $e_{g}$) band for each volume.
The positive and negative DOSs correspond to majority
and minority spin channels, respectively.
At the equilibrium volume, the insulating HS state is well-established,
while, at the LS stabilized volume, the half-metallic LS state is realized.
Results are in the GGA+$U$ scheme with $U$=2 eV.
}
\label{fsm_dos}
\end{figure}

CoCl$_{2}$ contrasts well with CoO, which is one of the most studied
Mott insulator governed by $U$ vs. $W$ physics.
In CoO, strong bonding between Co-$3d$ and O-$2p$
gives resilience to the system, and so resulting compressibility is small.
Consequently, the spin-state transition pressure is as high as 90 GPa,
which is about three times higher than that of CoCl$_{2}$.
Moreover, several structural transitions exist before reaching
the transition pressure \cite{cohen, guo, rueff, wdowik, zhang}.
Also the large total magnetic moment of CoO, 3.8 - 3.98$\mu_{B}$,
reflects the large orbital magnetic moment of Co$^{2+}$,
which hampers the accurate description of the electronic structure
of the system \cite{roth,jauch,norman}.

Figure \ref{fsm_dos}(a) provides the FSM calculational results of CoCl$_{2}$
for two different volumes,
the equilibrium (V/V$_{0}$=1.0) and V/V$_{0}$=0.72.
For V/V$_{0}$=0.72, the LS state is stabilized over
the HS state with $\Delta E$ of $\sim300$ meV. Related partial
density of states (PDOS) of Co-$3d$ band for each case is shown
in Fig.~\ref{fsm_dos}(b).  At the equilibrium volume,
the HS (t$_{2g}^{5}e_{g}^{2}$) state
with large band gap is obtained, while, for V/V$_{0}$=0.72,
the LS (t$_{2g}^{6}e_{g}^{1}$) metallic state is obtained,
which indicates that the spin-state transition is accompanied by the
insulator-to-metal transition.
Contrary to Mott systems that show highly increased bandwidth
upon pressure, the overall bandwidth of Co-$d$ in CoCl$_{2}$ does
not vary much, which reflects the restricted effect of $W$
in this system \cite{suppl}.
Interestingly, in the LS state,
the Fermi level cuts the spin-up $e_{g}$ band only,
so as to produce the half-metallic nature.
This point will be discussed more below.

In the case of Mott-type systems,
the simple DFT+$U$ (DFT: density functional theory) scheme
fails to describe the insulator-to-metal transition.
For example, for CoO, the DFT+$U$ approach reproduced
the observed pressure-induced HS to LS transition successfully,
but failed to manifest the insulator-to-metal transition \cite{zhang}.
The same failure occurred for MnO, too \cite{kasinathan}.
Kune\v{s} \emph{et al.} \cite{kunes08} have shown that
the dynamical treatment is necessary to explain
the spin-state transition in Mott systems.
In contrast, in the case of CoCl$_{2}$,
$U/W$ physics is not pronounced,
and so the description of insulator-to-metal transition
is possible in terms of the DFT+$U$.
The success in the description of CoCl$_{2}$ within DFT+$U$ scheme, in turn,
implies that the system is governed by $J$ vs. $\Delta_{CF}$ physics
\cite{suppl,calc}.

\begin{table}[b]
\centering
\caption{Critical volume and pressure at the spin-state transition
in CoCl$_{2}$ for each $U$ value.
P$_{E}$ is obtained from internal energy crossover,
and P$_{H}$ is obtained from the Maxwell construction
of enthalpy for the HS and LS states.
}
\begin{tabular}{ C{1.6cm}|C{1.6cm} C{1.6cm} C{1.6cm} C{1.6cm} }
\hline\hline
    $U$ (eV)         & 0    & 1.0    & 2.0   & 3.0   \\
\hline
   V/V$_{0}$         & 0.87 & 0.81 & 0.77 & 0.73 \\
   P$_{E}$ (GPa)     & 13.4 & 22.1 & 33.3 & 41.6 \\
   P$_{H}$ (GPa)     &  9.7 & 16.2 & 27.0 & 38.5 \\
\hline\hline
\end{tabular}
\label{U}
\end{table}

\begin{figure}[t]
\includegraphics[angle=270,width=76mm]{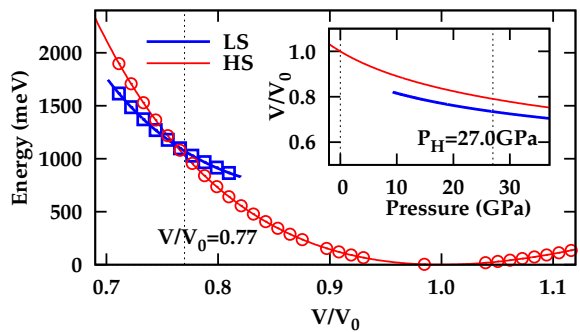}
\caption{(Color online) Energy ($E$) vs. volume curve
for CoCl$_{2}$ shows
the spin-state transition at V/V$_{0}$=0.77.
(inset) Volume vs. pressure curve obtained
from enthalpy $H=E+pV$. Abrupt volume collapse can be found
at the transition pressure denoted by the vertical dotted line.
Results are in the GGA+$U$ with $U$=2 eV.
}
\label{vol_press}
\end{figure}

 In Fig.~\ref{vol_press}, we have shown the energy-volume curves
for both HS and LS phases. The spin-state
transition is found at around V/V$_{0}$=0.77 for $U$=2eV.
In the inset, the volume vs. pressure curve is also shown.
Transition pressure (P$_{H}$) obtained from the Maxwell
construction is marked with vertical dotted line at 27.0 GPa,
which fits well with the experimental value of 30 GPa \cite{hernandez}.
The transition pressure is known to increase with $U$ \cite{zhang,kasinathan},
and we can find the same tendency for CoCl$_{2}$ too
(See Table~\ref{U}).
Appropriate $U$ value can be determined by comparison with the
experiment \cite{hernandez}, which agrees with our
calculation on the occurrence of insulator-to-metal transition.

Noteworthy is that, as in the case of Mott-type transition in MnO
and Fe$_{2}$O$_{3}$,
the abrupt volume collapse as large as 7.2$\%$ occurs
at the position of the spin-state transition
(See inset in Fig.~\ref{vol_press}) \cite{yoo,pasternak,badro}.
According to Pasternak \emph{et al.}\cite{pasternak}, the Mott transition
does not lead to a volume collapse at the pressure-driven transition.
The volume collapse manifested in CoCl$_{2}$ can be a good
complementary example suggesting the spin-state transition as a source of
volume collapse upon pressure.

The transition behavior of CoCl$_{2}$ is totally different
from that of Mott-type systems, such as  CoO and MnO.
While the latter show continuous transition
with the change of population in e$_{g}$ and t$_{2g}$ orbitals
for some pressure range \cite{cohen, huang, kunes08, mattila},
the former shows the sudden switch from HS to LS phase
due to its intrinsic metastable character of Co$^{2+}$
(Fig.~\ref{fsm_dos}(a)).
 Also the transition character in Co$^{2+}$ systems is totally
different from that in Co$^{3+}$ systems, for which the thermal excitation
promotes the population change of e$_{g}$ and t$_{2g}$ bands \cite{vogt,vanko}.

\begin{figure}[t]
\includegraphics[angle=270,width=76mm]{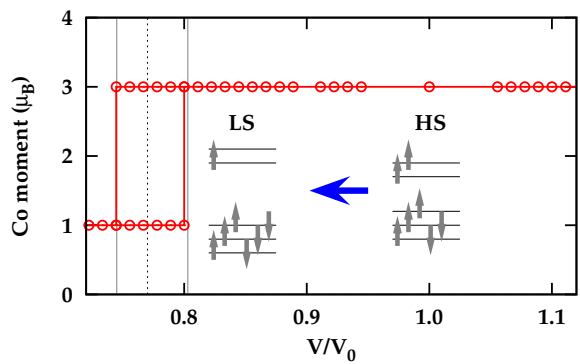}
\caption{(Color online)
Spin magnetic moment vs. volume curve
for CoCl$_{2}$ shows hysteresis behavior between HS and LS states.
The vertical dotted line corresponds to V/V$_{0}$=0.77,
corresponding to the internal energy crossover volume.
Schematic diagrams of corresponding HS and LS states are also shown.
}
\label{spin_state}
\end{figure}

The evolution of the magnetic moment as a function of volume
is described in Fig.~\ref{spin_state}.
One can find the hysteresis behavior as the HS phase is turned
to the LS phase and \emph{vice versa} between V/V$_{0}$=0.75 and 0.80.
 Since many HS Co$^{2+}$ systems have sizable orbital moment
due to incomplete quenching, the total moment change in some systems
can be larger than 2$\mu_{B}$ at the spin-state transition.

In CoCl$_{2}$, as the spin-state changes from HS to LS,
the magnetic structure is also found to change
from antiferromagnetic (AFM) to ferromagnetic (FM).
The HS AFM structure is
described as FM layers coupled antiferromagnetically along the hexagonal
$c$-axis \cite{hutchings,wilkinson}.
We compared total energies of AFM and FM CoCl$_{2}$ for different volumes,
and found that, at V/V$_{0}$=1.0,
the AFM state is favored by 5 meV/f.u., while, at V/V$_{0}$=0.72,
the FM state is favored by 15 meV/f.u.
Moreover, as shown in Fig.~\ref{fsm_dos}(b), the half-metallic state
emerges in the FM state at V/V$_{0}$=0.72, which
suggests that the double exchange mechanism becomes prevailing due to
the conducting $e_{g}$ electrons in the metallic LS phase.

In conclusion, we have confirmed the generality of the metastable LS state
in the octahedrally coordinated Co$^{2+}$ systems.
For CoCl$_{2}$, as a prototypical Co$^{2+}$ system,
we have demonstrated the pressure-induced spin-state transition,
which is governed by $J$ vs. $\Delta_{CF}$ physics.
Due to its high compressibility,
$\Delta_{CF}$ easily overturns $J$ at around 27 GPa,
and the first order spin-state transition occurs from HS to LS
with substantial volume collapse in company with the insulator-to-metal
and AFM to half-metallic FM transitions.
Since the Mott physics can be excluded,
we can argue that the spin-state transition and relevant behaviors
found in CoCl$_{2}$ are general features of a system with
$J$ vs. $\Delta_{CF}$ physics.

We would like to thank H. C. Choi and G. Lee for helpful discussions.
This work was supported by the NRF (No.2009-0079947),
and by the KISTI supercomputing center (No. KSC-2012-C2-27).


\begin{thebibliography}{99}
\bibitem{calc}
    Inclusion of the SOC does not change the overall
	electronic structure of the system.
	The orbital moment of 0.2 $\mu_{B}$ is obtained
	at the equilibrium volume, and 0.03 $\mu_{B}$
	at V/V$_{0}$=0.72.
    We also checked the effect of trigonal distortion of the system
	by changing $c/a$ ratio of CoCl$_{2}$ at each volume.
	The effect was negligible
	in view of the energy scale of the spin-state transition.
\end{thebibliography}
\end{document}